\documentclass[conference]{IEEEtran}
\IEEEoverridecommandlockouts
\usepackage{cite}
\usepackage{enumitem}
\usepackage{amsmath,amssymb,amsfonts}
\usepackage{algorithmic}
\usepackage{graphicx}
\usepackage{textcomp}
\usepackage{color, soul}
\usepackage[dvipsnames,table,xcdraw]{xcolor}
\usepackage{bm}
\usepackage{subfigure}
\usepackage{siunitx}
\usepackage{hyperref}
\usepackage[nocomma]{optidef}
\def\BibTeX{{\rm B\kern-.05em{\sc i\kern-.025em b}\kern-.08em
    T\kern-.1667em\lower.7ex\hbox{E}\kern-.125emX}}

\begin{document}

\title{CHIMERA: A Hybrid Estimation Approach to Limit the Effects of False Data Injection Attacks}% on Electric Power Grid}

% \title{Conference Paper Title*\\

%\author{Authors names are omitted for double-blind review}
\author{
\IEEEauthorblockN{\textbf{Xiaorui Liu}\IEEEauthorrefmark{1}$^{\ddag}$, \textbf{Yaodan Hu}\IEEEauthorrefmark{2}$^{\ddag}$, \textbf{Charalambos Konstantinou}\IEEEauthorrefmark{3}, \textbf{Yier Jin}\IEEEauthorrefmark{2}}
\IEEEauthorblockA{
\IEEEauthorblockA{\IEEEauthorrefmark{1}FAMU-FSU College of Engineering, Center for Advanced Power Systems, Florida State University\\
\IEEEauthorrefmark{2}Department of Electrical and Computer Engineering, University of Florida\\
\IEEEauthorrefmark{3}CEMSE Division, King Abdullah University of Science and Technology (KAUST)}
E-mail: xliu9@fsu.edu, cindy.hu@ufl.edu, charalambos.konstantinou@kaust.edu.sa, yier.jin@ece.ufl.edu}\thanks{$^{\ddag}$The first two authors contributed equally to this work. \protect 

This work is supported in part by Cyber Florida under Collaborative Seed Award \#3910-1011-00-A.}
}

\IEEEaftertitletext{\vspace{-1.4\baselineskip}}
\maketitle

\begin{abstract}
The reliable operation of power grid is supported by energy management systems (EMS) that provide  monitoring and control functionalities. Contingency analysis is a critical application of EMS to evaluate the impacts of outages and prepare for system failures. 
However, false data injection attacks (FDIAs) have demonstrated the possibility of compromising sensor measurements and  falsifying the estimated power system states. As a result, FDIAs may mislead system operations and other EMS applications including contingency analysis and optimal power flow. In this paper, we assess the effect of FDIAs and demonstrate that such attacks can affect the resulted number of contingencies. In order to mitigate the FDIA impact, we propose CHIMERA, a hybrid attack-resilient state estimation approach that integrates model-based and data-driven methods. CHIMERA combines the physical grid information with a Long Short Term Memory (LSTM)-based deep learning model by considering a static loss of weighted least square errors and a dynamic loss of the difference between the temporal variations of the actual and the estimated active power.  
Our simulation experiments based on the load data from New York state demonstrate that CHIMERA can effectively mitigate {91.74}\% of the cases in which FDIAs can maliciously modify the contingencies.

%the errors of the numbers of contingencies caused by attacks. 

% Although the integration of information and operational technology (IT/OT) aids the situational awareness of the power grid, the vulnerabilities of IT/OT devices introduce opportunities for adversaries to perform false data injection attacks (FDIAs) by compromising sensor measurements. 

% To maintain reliable operations of the electric power systems (EPS) against any sudden disturbance, an energy management system (EMS) is utilized by grid operators to monitor and optimize the EPS performance.
% Although the integration of the information and operational technology aids the situational awareness, the attacker could perform false data injection attacks (FDIAs) against system operations by compromising sensor measurements. The malicious measurements utilized in the state estimation may mislead the system operation and other EMS applications based on the state estimation results (e.g.,contingency analysis, power flow computing and etc). Contingency analysis is a critical application in EMS to evaluate the impacts of outages on the EPS. The system operators mainly rely on the results to identify the violations and be well-prepared for the potential system failure.

\end{abstract}


 %$\blfootnote{$^{\ddag}$The first two authors contributed equally to this work.\\This work is supported in part by Cyber Florida under Collaborative Seed Award \#3910-1011-00-A.}$

\vspace{-1mm}
\section{Introduction} \label{s:1}
\vspace{-1mm}

In electric power grids, energy management systems (EMS) provide situational awareness and assist the decision-making. EMS encompasses hardware/field components at geographically dispersed locations and telecommunications systems, as well as software applications at utility control centers, e.g., state estimation and contingency analysis.  Specifically, the network topology processor within EMS utilizes  %circuit 
breaker status and acquired data from telemetry  %and communication 
devices to update the power system model. The collected measurements and the updated system model facilitate the state estimator to determine the current system states. The estimated results are required by other EMS applications such as contingency analysis and optimal load flow algorithms. Thus, the accuracy %of all other 
EMS applications depends on the results of state estimation.

% Although the high integration of the information and operational technology (IT/OT) electronic devices brings the benefits of monitoring and control of industrial equipment, the existing vulnerabilities of these devices make the system more vulnerable to be attacked. 

As part of state estimation routines, bad data detection (BDD) units are used to identify anomalous measurements. However, it has been shown that false data injection attacks (FDIAs) can bypass BDD  \cite{liu2011false}. %The state estimation in multi-phase and unbalanced distribution systems is also vulnerable to FDIAs ~\cite{zhuang2019false}. 
Undetectable FDIAs under the situation of sensor failures could even worsen the estimation performance \cite{lu2020false}. In addition, the conditions of the 2015 attack on the Ukrainian  grid, demonstrated that the threat model of FDIAs could result in massive blackouts \cite{liang20162015}.

Contingency analysis is one of the core applications in EMS which evaluates the impact of the planned or unplanned problems that occur in the electric grid such as scheduled maintenance and component failures. Components refer to generators, transmission lines, transformers, circuit breakers, etc. According to the North American Electric Reliability Corporation (NERC), the fundamental criterion of $N-1$ ({where $N$ refers to the total number of components}) requires that the power system is able to withstand the disruption of one component outage \cite{NERC}. Contingency scenarios can be extended to $N-k$, which refers to a number of $k$ component failures. Grid operators rely on contingency analysis to recognize system overload conditions, rank the severity of the overloaded components, and isolate them if necessary to prevent cascading failures. % by comparing the calculated results with the threshold of operation limits such as power flow constraints and thermal constraints of transmission lines. If such limits are exceeded after a contingency event,  operators could isolate the overloaded components to prevent cascading failures. 
However, the reliability of contingency algorithms cannot be guaranteed when the system is under FDIAs \cite{kang2018false}. 

%Thus, a qualification is required to examine the impacts of FDIAs on contingency analysis in case the system operators is misled and a cascading failure appears in the electric power systems.  

% The authors in \cite{burada2016contingency} evaluate the severity of different contingencies by comparing the voltage change at buses and active power flow change at lines. 

%Talk about the hybrid detection here.. 

To detect the FDIAs, two major detection approaches are considered \cite{musleh2019survey}, model-based and data-driven methods. Model-based methods leverage system physics and data  (e.g., the grid topology and lines admittance) to estimate states with methods such as recursive weighted least square and Kalman filters \cite{sreenath2017recursive, kurt2018real}. %However, these conventional methods fail to identify malicious data injected by FDIAs. 
In order to determine whether or not an attack occurs, different tests are applied to the estimation results such as the large normalized residual~\cite{chu2020unobservable}, and the cumulative sum test \cite{kurt2018real}.
% Although the existing research on model-based techniques aiming to develop BDD methods resilient against FDIAs, 
However, such methods are typically computationally expensive in terms of processing time and scalability \cite{li2018pama}. On the other hand, despite the benefits of data-based approaches in terms of short execution times \cite{sayghe2020survey}, such techniques require a large set of training data to achieve good performance. 
% In the scenarios of network topology reconfigurations, such methods require retraining to ensure that learning models stay accurate. 
In addition, the rise of learning-based schemes in many applications is accompanied with important security challenges: it creates an incentive among adversaries to exploit potential vulnerabilities of the algorithms \cite{liu2019adversarial, saygheevasion}. 
{Recent works illustrated that combining the physics- with data- based models provides several advantages, especially in terms of security, as they tightly confine the solution scope and limit the capability of the adversarial examples \cite{anubi2019enhanced, wang2019physics}.}
% For instance, it is demonstrated solely artificial neural network-based estimation algorithms can be maliciously compromised \cite{aman2013energy}. 

%{Recent works illustrated that combining the physics-based models with data-based approaches provides several advantages, and specifically in terms of security, it tightly confines the solution scope and limits the capability of the adversarial examples \cite{anubi2019enhanced, wang2019physics, huang2019hybrid}. For example, in order to have robust state estimation results, the authors in \cite{wang2019physics} introduce a physics-guided deep learning model for time-series power system state estimation which uses the sequential measurements as inputs to reduce the high dependency of the current measurements. Moreover, a hybrid method is utilized in \cite{huang2019hybrid} to enhance the reliability of the state estimation results, where the data-driven estimator is combined with a topology identification method to track system states and the model-based estimator is used for filtering the noises and gross errors from the measurements.}

%  melacci2020can

%in this paper, we integrate the system-level information into the deep learning models and develop a physics-guided neural network which can predict the states with resilience to FDIA. 

% In this paper, we investigate the possibility of embedding system-level information, such as balanced power flow equations, into the neural networks to enhance the resilience of the models to adversarial perturbations. 

{In this paper, we study the impacts of FDIAs  and propose a hybrid, model-based and data-driven,  attack-resilient state estimator to mitigate the attack impact on the contingency analysis results.  To the best of our knowledge, this paper is the first study to propose a hybrid estimation approach on how to mitigate the effect of FDIAs on contingency analysis. } Our contributions are summarized as follows:
%\vspace{-1mm}
\begin{itemize}[leftmargin=*]
    \item We formulate an attack model to bypass state estimation BDD and cause, via FDIAs, non-critical transmission lines, i.e., lines not included in the contingency screening, to surpass their power flow limits. We show that the FDIAs impact %on contingency analysis %based on this attack model 
    can effectively distort the number of system contingencies. 
    \item To mitigate the attack impact, we propose CHIMERA\footnote{CHIMERA, according to Greek mythology, was a monstrous fire-breathing \emph{hybrid} creature composed of several different animals.}, a hybrid attack-resilient state estimator. i.e., a physics-informed estimator constructed based on Long Short Term Memory (LSTM) networks. It embeds the grid observation model of power flow equations into neural networks. We exploit the static and dynamic features of the  observation model to construct spatial-temporal correlations among measurements, and limit how FDIAs against state estimation can affect subsequent EMS contingency results.
    \item We conduct simulation experiments based on load  data  from  New  York  state. The results demonstrate that CHIMERA can effectively mitigate 91.74\% of the attack cases in which FDIAs can  maliciously modify the contingency results.
\end{itemize}
 %Our simulation experiments demonstrate that CHIMERA is resilient to FDIAs and the impacts of FDIAs on contingency analysis can be mitigated.

The rest of this paper is as follows: Section \ref{s:2} provides background information. Section \ref{s:3} discusses our attack model, and Section \ref{s:4} presents the mitigation strategy. Experiments are shown in Section \ref{s:5}. Section \ref{s:6} draws concluding remarks.
\vspace{-2mm}
\section{Background} \label{s:2}
\vspace{-1mm}
%Since this paper combines concepts from various fields, we have gathered short descriptions of state estimation, contingency analysis, and long short term memory concepts to give some completeness and improve readability. 

\subsection{State estimation}
% State estimation refers to a process which estimates the current power system state by utilizing the collected measurements (e.g., supervisory control and data acquisitions (SCADA) system, phase measurement units (PMUs), meters, and etc.). 
\vspace{-1mm}
In the nonlinear (AC) state estimation of power systems, state variables are determined by phase angles ($\theta$) and voltage magnitudes ($V$). % which reflects the network operating condition. 
For a system with $n$ buses, the states are $\mathbf{x}=\left[\theta_{2}, \theta_{3}, \ldots, \theta_{n}, V_{1}, \ldots, V_{n}\right]^{T}$, where $\theta_{1}=0$ is the reference angle. To maintain full observability of the system, $m \ge n$ measurements are required. Measurements ($\mathbf{z}$) %is defined as $\mathbf{z}=\left[P_{1}, P_{2}, \ldots, P_{m}, Q_{1}, \ldots, Q_{m}\right]^{T}$ which 
typically include active power ($P$) and reactive  power ($Q$) measurements. The relationship between %system 
states and  acquired measurements, with $\mathbf{e}$ being a vector of noises, 
 is as follows: %given in Eq. \eqref{Eq:measurement}. 

\vspace{-2mm}
\begin{equation}
\mathbf{z}=\mathbf{{h}}(\mathbf{x})+\mathbf{e}
\label{Eq:measurement}
\vspace{-1mm}
\end{equation}
% \begin{equation}
% \begin{aligned}
% P_{i} &=V_{i} \sum_{j \in N_{i}} V_{j}\left(G_{i j} \cos \theta_{i j}+B_{i j} \sin \theta_{i j}\right)\\
% Q_{i} &=V_{i} \sum_{j \in N_{i}} V_{j}\left(G_{i j} \sin \theta_{i j}-B_{i j} \cos \theta_{i j}\right)
% \label{Eq:pqi}
% \end{aligned}
% \end{equation}

% \begin{equation}
% \begin{aligned}
% P_{i j}=V_{i}^{2}\left(g_{s i}+g_{i j}\right)-V_{i} V_{j}\left(g_{i j} \cos \theta_{i j}+b_{i j} \sin \theta_{i j}\right)\\
% Q_{i j}=-V_{i}^{2}\left(b_{s i}+b_{i j}\right)-V_{i} V_{j}\left(g_{i j} \sin \theta_{i j}-b_{i j} \cos \theta_{i j}\right)
% \label{Eq:pqij}
% \end{aligned}
% \end{equation} 
 
%%%%%%%%%%%%%%%%%%%%%%%%%%%%%%%%%%%%%%%%%%%%%%%%%%%%%%%%%%%%%%%%%%%%%%%% 
% To solve Eq.(\ref{Eq:measurement}) and obtain the optimal estimated state $\hat{\mathbf{x}}$, a WLS method is commonly utilized by minimizing the residual $J(\mathbf{x})$ as shown in Eq.(\ref{Eq:jx}).  

State estimation is widely solved via iterative techniques such as the weighted least square method \cite{konstantinou2016case}, in which the accuracy of the estimated variables ${\mathbf{x}}$ is
calculated via the Euclidean norm of the residual $\lVert \mathbf{z} - \mathbf{h}(\hat{\mathbf{x}}) \rVert$. For example, the estimated states $\mathbf{\hat{x}}$ can be obtained through optimization  of $J(\mathbf{\hat{x}})$ in Eq. \eqref{Eq:jx}, where $\mathbf{W}=\operatorname{diag}\left\{\sigma_{1}^{-2}, \sigma_{2}^{-2}, \cdots, \sigma_{m}^{-2}\right\}$.% composes all the weights $\sigma^{-2}$ of the measurements.  
There are different approaches to solve Eq. \eqref{Eq:jx}; one such method is via iteratively solving Eq. \eqref{Eq:x}. To detect whether or not the state estimation is disturbed by the random noises or attacks, BDD compares the objective function $J(\mathbf{\hat{x}})$ with a normalized threshold $\tau$. If $J(\mathbf{\hat{x}}) < \tau$, no bad data is detected. % Otherwise, the system operator will be notified for subsequent calculations to reveal the bad data if the condition is not satisfied. 
%\vspace{-1mm}
\begin{equation}
\min _{\hat{x}} J(\mathbf{\hat{x}})=(\mathbf{z}-\mathbf{h} (\mathbf{{\hat{x}}}))^{T} \mathbf{W}(\mathbf{z}-\mathbf{{h} ({\hat{x}}))}
\label{Eq:jx}
\end{equation}
\vspace{-1mm}
\begin{equation}
\mathbf{H}^{T}_{k} \mathbf{W} \mathbf{H}_{k} \Delta {\mathbf{\hat{x}_{k}}} = \mathbf{H}^{T}_{k} \mathbf{W} [\mathbf{z}-\mathbf{h}(\mathbf{\hat{x}_{k}})]
\label{Eq:x}
\vspace{-1mm}
\end{equation}

To reduce the computational overhead, linear (DC) state estimation is often adopted which assumes that transmission line resistances are negligible, voltage magnitudes are 1 per unit, and the differences in voltage angles between buses are small. Thus, the observation model can be linearized: %as follows:
\vspace{-1mm}
\begin{equation}
        P_i = \sum\nolimits_{j \in N_i}{\mathbf{B}_{ij}}(\theta_i - \theta_j), 
\label{eq:pi}
\vspace{-1mm}
\end{equation}
and in matrix form $ \mathbf{P} =\mathbf{ H} \bm{\theta} $, in which $\mathbf{P}$ and $\bm{\theta}$ are the vectors of the active power measurements and the voltage angles of the buses, respectively. $\mathbf{H}$ is the measurement Jacobian matrix derived from the susceptance matrix $\mathbf{B}$. With the approximations, the accuracy of the estimation is decreased while the computation overhead is reduced. The states $\bm{\theta}$ can be estimated with the following equation:
\vspace{-1mm}
\begin{equation}\label{Eq:DC_se}
    {\tilde{\bm{\theta}}} = (\mathbf{H}^T \mathbf{H})^{-1} \mathbf{H}^T \mathbf{P}
    \vspace{-1mm}
\end{equation}

\subsection{Contingency Analysis}\label{sec:CA}

% Contingency analysis simulates and evaluates the impact of the planned or unplanned problems that occur in the EPS such as scheduled maintenance and system component failures. The components refer to the generator, transmission line, transformer, circuit breaker, and etc. To assess the performance of different systems, the specific contingency scenario has been sorted into different approaches. The fundamental criterion is $N-1$ which is enforced for each utility's power system by The North American Electric Reliability Corporation (NERC) \cite{NERC}. It requires that the system is able to withstand the disruption of any one component outage. Moreover,

Contingency analysis simulates the effects of contingency/outage scenarios and calculates the overload conditions. However, the computational cost of such ``what-if'' scenarios is unrealistic for large-scale and complex power systems. The computational overhead is proportional to $N!/[k!(N-k)!]$ for $N-k$ contingencies. Due to the low probability
of $N-3$ contingencies occurring in different transmission lines in real-world \cite{chien2007automation}, research works typically focus on $N-1$ and $N-2$ scenarios \cite{liu2020deep}. %Other than the base case $N-1$, $N-2$ criteria is applied to the higher power supply reliability area which indicates that the system could tolerate the outage of two components simultaneously. 
In order to find all power flow constraint violations under $N-1$ and $N-2$ scenarios, the linear power flow approximation is typically utilized \cite{8298523}. 
Following such approach, in this work, the power flows are calculated by $\mathbf{f}=\mathbf{Y}\mathbf{M}^{T}\bm{\theta}$, where $\mathbf{Y}$ is the branch susceptance matrix, $\mathbf{M}$ is the connection matrix, and $\bm{\theta}$ is the vector of voltage phase angles. Additionally,  $\mathbf{f}$ is used to calculate the line outage distribution factors (LODFs). LODFs determine the power flow impact on the remaining lines when one or more line outages are observed in the system. The formulation of single and double outages can be found in \cite{liu2019reinforcement}.

\vspace{-1mm}
\section{Attack Model} \label{s:3}
\vspace{-1mm}

%  The collected system information from SCADA (e.g., power injection, voltage magnitude, circuit breaker statues and etc) is utilized in the state estimation process to compute the optimal estimated state. Based on the estimated state variables, the optimal power flow of the system is determined and then used in the contingency analysis to assess the impacts of component failures on the entire power grid. The contingency analysis informs the potential vulnerabilities of the grid which allow an automatic generation control scheme or the system operators to adjust and fortify a reliable operation.

% In this section, we provide the threat model of the work. FDIAs have been traditionally demonstrated on how compromise the state estimation process against system operation \cite{musleh2019survey}. Our proposed attack model assumes  that the attacker is not solely target to falsify state estimation but also manipulate the contingency analysis results. A mathematical formulation of the attack model is provided in which the  attacker could overload the most vulnerable line found in contingency analysis through FDIAs.

% alter the resultant number of contingencies 

\subsection{Threat Model}
FDIAs have been traditionally demonstrated on how to compromise the state estimation \cite{musleh2019survey}. In this paper, we assume that the attacker does not solely target to falsify the state estimation but also to manipulate the contingency analysis results \cite{9351954}. We consider an attacker who can exploit the configuration of a power system to launch FDIAs by manipulating the sensor measurements while bypassing BDD. Moreover, the attacker targets those measurements which could distort the number of contingencies. The assumptions of the threat model are as follow: % In our adversarial model, the attacker aims to compromise certain measurements with the following assumptions: 
\begin{itemize}[leftmargin=*]
\item The attacker has full observation of the topology and configurations of the system, i.e., the attackers could construct the Jacobian matrix $\mathbf{H}$. Such data can be obtained through public information or signal reconstruction  \cite{keliris2019open,  kim2014subspace}.  
\item The attacker is aware of the specifics of the state estimation process, either model-based or data-driven, in order to carefully craft FDIAs to bypass BDD routines. 

% \item The attacker is aware of the existence of the BDD function ($J\mathbf{(x)}$) as well as the threshold ($\tau$).
\item The attacker has access to the real-time measurements of deployed grid sensors, e.g., via eavesdropping on the communication links. However, due to the limited physical access or the protection of certain meters, the attacker can compromise a limited number of measurements \cite{konstantinou2016case}. 

\item The attacker could perform contingency analysis based on the estimated results and the power flow constraints of each line required to ensure an overload condition \cite{liu2020deep}. 
\end{itemize}
 
\vspace{-2mm}
\subsection{Mathematical Formulation}

Despite BDD mechanisms can detect  measurement anomalies, carefully crafted FDIAs can bypass such algorithms. Consider the malicious vector $\mathbf{a}$ injected into measurements $\mathbf{z}$, then the compromised vector can be represented as $\mathbf{z_{a}}=\mathbf{z}+\mathbf{a}$. The  attacked estimated state variables can be written as $\mathbf{\hat{x}_{a}}= \mathbf{\hat{x}}+\mathbf{c}$, where $\mathbf{c}$ is the vector of the injected and resulted error. A successful FDIA undetected by the residual-based BDD, as shown in Eq. \eqref{Eq:fdi}, can be formed when $\mathbf{a} = \mathbf{Hc}$, i.e., if $\mathbf{a}$ is a linear combination of $\mathbf{H}$, for the arbitrary vector $\mathbf{c}$. 

\vspace{-2mm}
\begin{equation}
\begin{aligned}
\lVert {r_{a}} \rVert &=
\left\|\mathbf{z_{a}}-\mathbf{H}( \mathbf{\hat{x}_{a}})\right\| \\
&=\left\|\mathbf{z}+\mathbf{a}-\mathbf{H}(\mathbf{\hat{x}}+\mathbf{c})\right\| \\
&=\left\|\mathbf{z}+ \mathbf{H}(\mathbf{\hat{x}}+\mathbf{c})-\mathbf{H}(\mathbf{\hat{x}})-\mathbf{H}(\mathbf{\hat{x}}+\mathbf{c})\right\| \\
&=\|\mathbf{z}-\mathbf{H}(\mathbf{\hat{x}})\| =\lVert r \rVert 
\end{aligned}
\vspace{-1mm}
\label{Eq:fdi}
\end{equation}

\begin{figure}[t]
 \centerline{\includegraphics[width=\columnwidth]{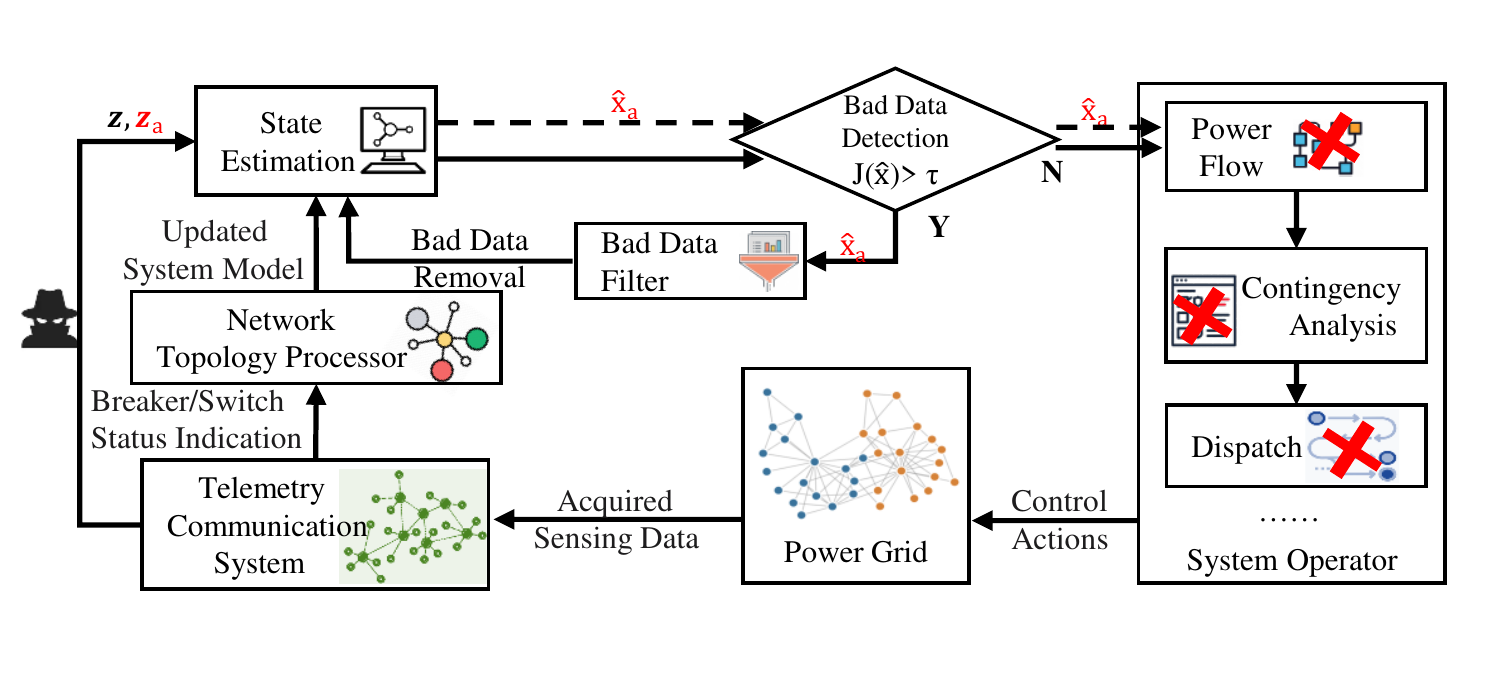}}
%  \centerline{\includegraphics[width=0.5\textwidth]{Fig/Fig1AttackModel.pdf}}
  \vspace{-5mm}
\caption{Illustration of the attack model (attacked variables in red color).}
\label{fig:attackmodel}
  \vspace{-4mm}
\end{figure}

Fig. \ref{fig:attackmodel} depicts the overall process of the attack model. The sensor measurements $\mathbf{z}$ are compromised by FDIAs represented by an attack vector $\mathbf{a}$. The estimated states $\mathbf{\hat{x}}$, as the output of the estimation process, will be altered under FDIAs. The BDD can detect and remove the significant errors as bad data, namely $J\mathbf{(\hat{x})}$ above the threshold $\tau$. Otherwise, if BDD is bypassed due to FDIAs, the malicious states will be processed to perform contingency analysis. Since the power flow computation, $\mathbf{f^a}$, is affected, the contingency %analysis 
results $\mathbf{f^a}^{\prime}$ will be inaccurate. As a result, system operators will be misled by the malicious contingency analysis output,  and thus, potential threats to the power system reliability may be posed.

Based on the assumptions of the attacker's capabilities and knowledge of the power system topology and data, the attack model is mathematically formulated as Eq.~\eqref{eq:eq1} - \eqref{eq:con6}, where the attacker's objective is to affect the results of contingency analysis by FDIAs. In order to achieve that, the attacker performs contingency analysis to obtain the power flows under contingency and find the most vulnerable line $i$ which has the smallest difference between its power flow under contingency ${{f_{i}^\mathbf{a}}^{\prime}}$,  and its power flow capacity $f^{limit}_i$. The targeted line will overload based on the maximization function with an optimal attack vector $\mathbf{a}$ through FDIAs, as shown in Eq.\eqref{eq:eq1}. An absolute value of ${{f_{i}^\mathbf{a}}^{\prime}}$ is used here to represent the overflow observed either with ${{f_{i}^\mathbf{a}}^{\prime}}>f^{limit}_i$ or $-{{f_{i}^\mathbf{a}}^{\prime}}>f^{limit}_i$.

In order for the FDIAs to be stealthy and not being detected, several constraints represented by Eq.~\eqref{eq:con1} - \eqref{eq:con6} should be satisfied. In practice, a safety margin $f_{m}$ in the line flow capacity is reserved to reduce the overload risk. Therefore, only the line with a power flow below the certain line flow capacity ${f^{limit}_i} - f_{m}$ will be considered, as described in \eqref{eq:con1}. Eq.\eqref{eq:con2} shows that the attacker can compromise certain measurement $\mathbf{z}$ to $\mathbf{z_a}$ by adding an attack vector $\mathbf{a}$. Accordingly, the estimated state variables $\mathbf{\hat{x}}$ will be deviated to $\mathbf{\hat{x}_a}$ in \eqref{eq:con3}. In order to maintain stealth and bypass the BDD, the injected error should guarantee that the residual $J(\mathbf{\hat{x}_a})$ is within the system threshold $\tau$, as depicted in \eqref{eq:con4}. Once the malicious state variables are utilized to perform power flow computations, the results $\mathbf{f^a}$ will be affected since the voltage phase angles $\bm{{\hat{\theta}_\mathbf{a}}}$ in \eqref{eq:con5} are part of the deviated estimated variables $\mathbf{\hat{x}}$. The factor $\lambda^\mathbf{a}$ to qualify the line overload condition, LODF, will be utilized to compute the power flows. Taking the compromised power flow equation at line $i$ (${f_{i}^\mathbf{a}}$), line $j$ (${f_{j}^\mathbf{a}}$) with the LODF, the power flow of line $i$ under FDIAs with line $j$ during a outage is derived as ${{f_{i}^\mathbf{a}}^{\prime}}$ in \eqref{eq:con6}.

\vspace{-6mm}
\begin{maxi!}|l|[2]                  
    {\mathbf{a}}                               
    {\mathop{argmin}\limits_{{|f^\mathbf{a}_i}^{\prime}|} 
f^{limit}_i - |{f^\mathbf{a}}^{\prime}_i|  \label{eq:eq1}}   
    {\label{eq:Example1}}             
    {}                                
    \addConstraint{ |{f^\mathbf{a}}^{\prime}_i|}{< {f^{limit}_i} - f_{m} \label{eq:con1}}  
    \addConstraint{\mathbf{z_{a}}}{=\mathbf{z}+\mathbf{a}  \label{eq:con2}}  
    \addConstraint{\mathbf{\hat{x}_a}}{=\mathbf{\hat{x}} + \mathbf{c}  \label{eq:con3}}  
    % \addConstraint{J({\mathbf{\hat{x}_a}})=(\mathbf{z}-\mathbf{{h}({\hat{x}_a}}))^{T} {W}(\mathbf{z}-\mathbf{{h} ({\hat{x}_a}}))}{<\tau  \label{eq:con4}} 
     \addConstraint{J({\mathbf{\hat{x}_a}})<\tau  \label{eq:con4}} 
    \addConstraint{\mathbf{{f^{a}}}}{=\mathbf{Y} \mathbf{M^{T}} \bm{{\hat{\theta}_\mathbf{a} }}  \label{eq:con5}}  
    \addConstraint{{{f_{i}^\mathbf{a}}}^{\prime}}{={\lambda_{i j}^\mathbf{a}} {f_{j}^\mathbf{a}} + {f_{i}^\mathbf{a}}  \label{eq:con6}}  
\end{maxi!}

\vspace{-2mm}
\section{CHIMERA: Hybrid Attack-Resilient  Estimator}\label{s:4}
%As presented in Fig.~\ref{fig:architecture}, 
In order to mitigate the impacts of FDIAs on the state estimation and the consequential contingency analysis, we propose CHIMERA, a hybrid attack-resilient state estimator. CHIMERA is an AC state estimator, which takes active and reactive power measurements as well as DC-estimated voltage angles as the input, and provides estimates of voltage magnitudes and angles of the buses. Given the attack model presented in Section \ref{s:3} and considering that a DC power flow model is typically used in grid operations \cite{musleh2019survey}, we build an AC hybrid estimator which is resilient to FDIAs affecting EMS routines. Despite a corrupted DC estimation output $\tilde{\bm{\theta}}_t$, CHIMERA provide accurate state predictions, by taking advantage of both the observation model Eq.~(\ref{Eq:measurement}) and an LSTM-based deep learning model. The LSTM network can capture the temporal correlations between data, and thus, the errors induced by the attacks can be corrected by the historical information. Moreover, since the observation model can confine the solution space with the physical constraints, we construct the loss function based on such a model.  % which are introduced in detail later.
In regards to the enhancement of the convergence speed and the estimation accuracy, we provide the DC estimation results ${\tilde{\bm{\theta}}}$ as the input of CHIMERA in addition to the power measurements ($\mathbf{P}$, $\mathbf{Q}$). Despite the limited accuracy of the DC estimation results due to the approximations, the DC estimated voltage angles can directly infer the scope of the true voltage angles. Thus, even in the presence of FDIAs, we include the DC estimated voltage angles in the input of CHIMERA because they can partially represent the states of the power grid. As a result, the input is formulated as $\mathbf{u}_t = [\mathbf{z}_t; \tilde{\bm{\theta}}_{t}]$, in which $\mathbf{z}_t$ and $\tilde{\bm{\theta}}_t$ are the vectors of the sensor measurements and the DC estimated states at time $t$, respectively. 

To capture the spatio-temporal correlations of the observation model in the presence of benign and malicious data,  the loss function of CHIMERA is composed of two parts: the static loss and the dynamic loss. In general, to regulate the accuracy of the estimated states, a typical way is to use the difference between the observed measurements and the derived measurements from the observation model Eq.~(\ref{Eq:measurement}) \cite{wang2019physics, liu2019adversarial}, which is defined as the static loss:
 \vspace{-1mm}
\begin{equation}\label{Eq:static}
    L_{static} = MSE({\mathbf{z}_t}, \mathbf{h}({\hat{\mathbf{x}}_t})),
\end{equation}
in which $\hat{\mathbf{x}}_t = [\hat{\bm{\theta}}_t;\hat{\mathbf{V}}_t]$ is the vector of estimated states from the model.  % and consists of the vector of voltage angles $\hat{\bm{\theta}}_t$ and magnitudes $\hat{\mathbf{V}}_t$.
$MSE(\mathbf{x}, \mathbf{y}) = (1/n) \sum_{i=1}^{n}\left(x_{i}-y_{i}\right)^{2}$ is the Mean Squared Error (MSE) between $\mathbf{x}$ and $\mathbf{y}$.  
Nevertheless, with only $L_{static}$, the LSTM network cannot totally mitigate the impacts of FDIAs, especially on contingency analysis.  
Although the structure of LSTM can utilize the temporal correlations of data implicitly, adversarial perturbations including FDIAs on such recurrent neural networks have been proven effective \cite{sayghe2020survey}. Therefore, as also shown in Section~\ref{s:5},  depending solely on the temporal correlations from LSTM is insufficient to defend against the attack proposed in Section~\ref{s:3}. To better describe the temporal correlations between data, we further exploit the consistency of the observation model in the time domain and explicitly augment the loss function with the dynamic loss, $L_{dynamic}$. The dynamic loss measures the distance between the expected and the actual variations of the measurements. Given Eq.~(\ref{eq:pi}), we have: 
 \vspace{-1mm}
\begin{align}
    \mathbf{P}_t - \mathbf{P}_{t-1} = \mathbf{H} (\hat{\bm{\theta}}_t - {\tilde{\bm{\theta}}}_{t-1}),
\end{align}
in which $\hat{\bm{\theta}}_t$ is the vector of the phase angles estimated by CHIMERA. Denote $\mathbf{\Delta {P}}_t = \mathbf{P}_t - \mathbf{P}_{t-1}$ and $\mathbf{\Delta \hat{{P}}}_t = \mathbf{H} (\hat{\bm{\theta}}_t - \tilde{\bm{\theta}}_{t-1})$.  Thus the dynamic loss is defined as: 
\begin{equation}\label{Eq:dynamic}
    L_{dynamic} = MSE(\mathbf{\Delta {P}}_t, \mathbf{\Delta \hat{{P}}}_t).
\end{equation}

\begin{figure}[t]
    \centering
    \includegraphics[width = 0.5\textwidth]{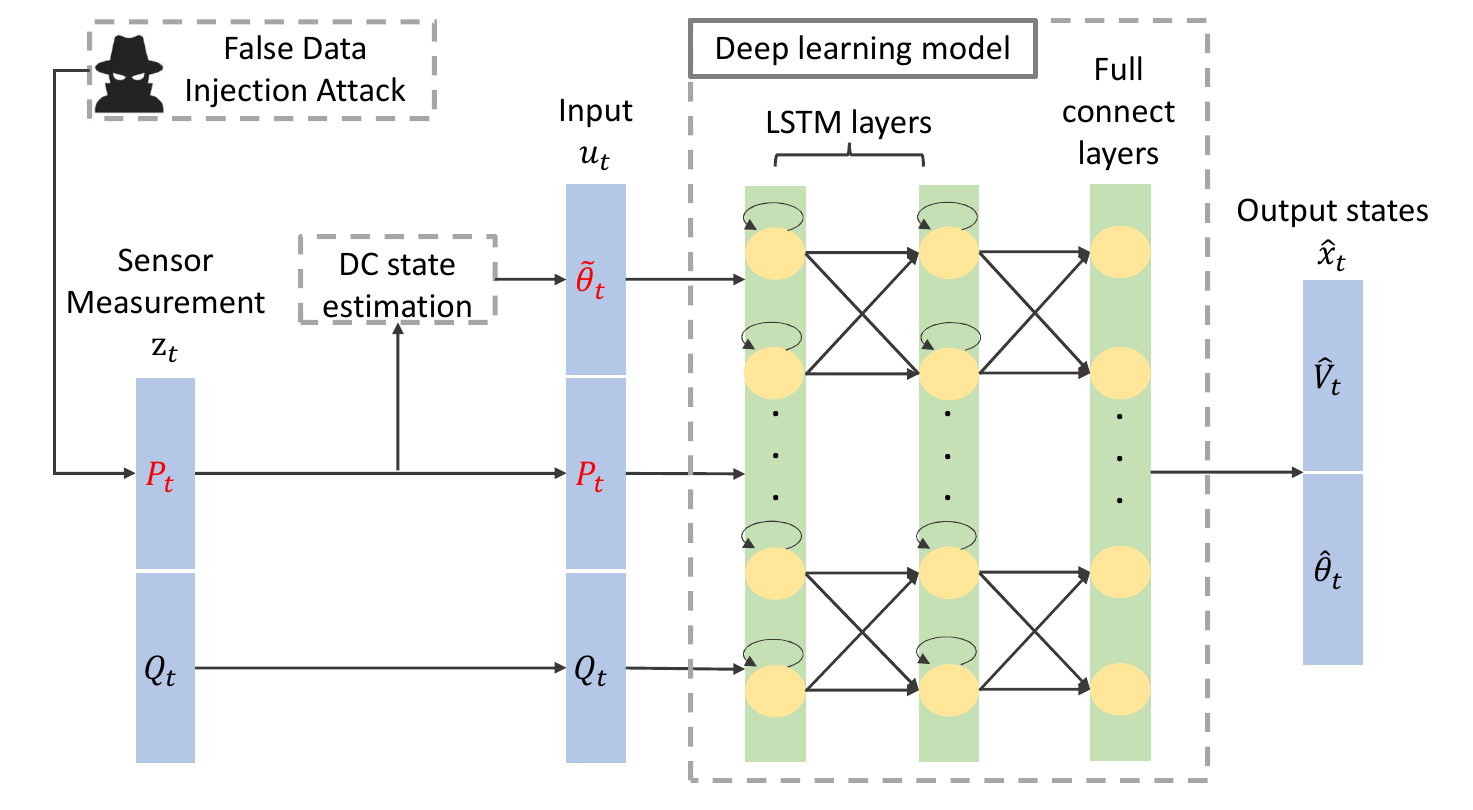}
    \vspace{-4mm}
    \caption{The architecture of CHIMERA. The  variables with red color are the ones that can be affected by FDIAs.}
    \label{fig:architecture}
    \vspace{-5mm}
\end{figure}

The static loss, $L_{static}$, guarantees that the observation model is satisfied at each epoch, and the dynamic loss, $L_{dynamic}$, enforces the temporal consistency between the estimated states and the system measurements. Given $L_{dynamic}$ and $L_{static}$, the loss function of CHIMERA is defined as:
%%\vspace{-1mm}
\begin{small}
\begin{equation}\label{Eq:loss}
    L = L_{static} + \gamma L_{dynamic},
\end{equation}
\end{small}
where $\gamma \leq 1$ is the weight to balance the two terms.
Compared with the loss function directly using the MSE between the estimated states and the true states, i.e., $L_0 = MSE(\hat{\mathbf{x}}, \mathbf{x})$ ($\mathbf{x}$ is the vector of the true states), the proposed loss function $L$ has several advantages. 
The true state $\mathbf{x}$ is not required in $L$. Note that solving $\mathbf{x}$ is non-trivial. The weighted least square method usually utilizes iterative methods to recursively minimize the residual $J(\mathbf{x})$ in Eq.~(\ref{Eq:jx}), which is often time-consuming,  especially when the grid size increases. 
Therefore, the utilization of $L_{static}$ and $L_{dynamic}$ can boost timing performance while estimating the true system states. Besides, as mentioned in Section~\ref{sec:CA}, the accuracy of contingency analysis heavily relies on the accuracy of the estimated variables. Since $L_{static}$ measures the difference between estimated and true power flows, $L_{static}$ can enhance the accuracy of contingency results by enforcing the consistency between the estimated and true power flows. {Unlike other approaches focusing on FDIA detection, CHIMERA is `fertilized' with the resilient estimation capability. This ensures that CHIMERA remains secure against other formulations of FDIAs because the attack impact will be restrained as long as the accuracy of the estimation process is guaranteed.}

The architecture of CHIMERA is depicted in Fig.~\ref{fig:architecture}. To avoid over-fitting, validation data is utilized to select the most suitable hyper-parameters for CHIMERA. We run CHIMERA with different configurations and select the one with the most accurate estimations on the validation data. The detailed configuration is explained as follows. CHIMERA is composed of two LSTM layers and a full connection layer. For each LSTM layer, the number of the features in the hidden state is 128 and the length of the sequence is 32. For the loss function, we set $\gamma = \num{1e-3}$. During the training phase, a batch of vectors $\mathbf{u}_t$ with batch size 32 are provided as input. The outputs from the output layer $\mathbf{\hat{x}}_t$ are then used to calculate the loss based on Eq.~(\ref{Eq:loss}). The weights and the biases of the model are updated with the gradient of $L$ by the Adam algorithm \cite{kingma2014adam}, through the back-propagation process. Due to the non-linearity of the observation model in Eq. \eqref{Eq:measurement}, there are many local minimums. To approach the global minimum of $L$, we train the model with two steps. We first train a coarse model with a large learning rate \num{1e-3} for 150 iterations. Then the model is fine-tuned for 500 iterations with small learning rates varying following a triangular cycle, which linearly increases from \num{1e-7} to \num{1e-4} and then decreases back to \num{1e-7}.

 \vspace{-2mm}
 \section{Evaluation and Simulation Results}\label{s:5}

\subsection{Experimental Setup}
\subsubsection{Dataset}
To examine the impacts of FDIAs on contingency analysis, we compare the number of contingencies and the overload conditions when the system is operated in normal conditions and under FDIAs. We conduct the experiment based on the IEEE 14-bus system and use synthetic data generated from the load data provided by the New York Independent System Operator (NYISO). We use NYISO load data from May 2020 containing the 5-min-interval active powers at each NY region, with 9030 epochs in total. The synthetic data is generated according to \cite{anubi2019enhanced}, and 
%The IEEE 14 bus system includes 5 generators, 11 loads, and 14 buses as shown in Fig.\ref{fig:IEEE14}. Each load bus of the IEEE 14 bus system corresponds to one of the 11 regions of NYISO (Fig.~\ref{fig:NYISO}). For example, the collected data from region $A$ of NYISO named "West" is utilized to simulate the power consumption on bus 2. The detailed mapping between the buses and the regions is presented in Table~\ref{tab:mapping}.
due to the unavailability of reactive power information, we generate the reactive power data by assuming a constant power factor of 0.8. 
%The active and reactive powers are normalized to the initial active and reactive powers of the IEEE 14 bus system, respectively. Besides, the generator ramp rates are set at the ratio of the derived total active/reactive powers to the initial total active/reactive powers of the IEEE 14 bus system. The grid measurements include the active/reactive powers at each bus and the states are the voltage magnitudes and angles of each bus. We leverage Matpower \cite{zimmerman2010matpower} to calculate the measurements and the states based on the derived load data using Newton’s method. 
%Similar to \cite{wang2019physics}, 
White Gaussian noises with means of 0 and standard deviations of 0.01 are added to the measurements. We regard the measurements and the states generated as the ground truth when evaluating the performance of the estimation model. When executing contingency analysis, we use the flow limits listed in \cite{venkatesh2003comparison}. %, and presented in Table~\ref{tab:flimit}.

\iffalse
\begin{table}[t]
    \centering
    \caption{The mapping between the IEEE 14 bus system and the NYISO control regions.}
    \begin{tabular}{|c|c|c|c|c|c|c|c|c|c|c|c|}
        \hline
         Zone & A & B &C & D & E & F & G & H & I & J & K\\
         \hline
         Bus & 2 & 3 & 4 & 5 & 6 & 9 & 10 & 11 & 12 & 13 & 14 \\
         \hline
    \end{tabular}
    \label{tab:mapping}
\end{table}

\begin{figure}[t]
\centering    
\subfigure[] { \label{fig:IEEE14}     
\includegraphics[width = 0.18\textwidth]{Fig/IEEE14.pdf}
 }
% \hfill
\subfigure[] { \label{fig:NYISO}     
\includegraphics[width = 0.25\textwidth]{Fig/nyiso_map.jpg}
}
\caption{ \subref{fig:IEEE14} IEEE 14 bus system, and \subref{fig:NYISO} New York Control Load Zones.} 
\label{fig:resultsattacks}  
\end{figure}
\fi

% \begin{figure}[!htbp]
%     \centering
%     \includegraphics[width = 0.42\textwidth]{Fig/IEEE14.pdf}
%     \caption{IEEE 14 bus system.}
%     \label{fig:IEEE14}
% \end{figure}

% \begin{figure}[!htbp]
%     \centering
%     \includegraphics[width = 0.47\textwidth]{Fig/NYISO.pdf}
%     \caption{New York Control Load Zones.}
%     \label{fig:NYISO}
% \end{figure}

\subsubsection{Deep Learning Models for Evaluation}
In addition to CHIMERA, we train two models for comparison purposes:  a Multilayer Perceptron (MLP) network and the model proposed in \cite{wang2019physics}. Since MLP induces limited computational overhead, it has been widely applied to the power grid \cite{mosbah2015multilayer}. In this paper, we train a MLP network as the performance baseline. The MLP is composed of three hidden layers with 128 neurons for the first two layers and 64 neurons for the last hidden layer. We use $L_0 = MSE(\hat{\mathbf{x}}, \mathbf{x})$ as the loss function of MLP. Therefore, no additional information or system dynamics are leveraged to defend against FDIAs. We refer to this model as the baseline MLP and use it to demonstrate the impacts of FDIAs when no defense is considered. Besides the baseline MLP, we also utilize for comparison the physics-guided deep learning network proposed in \cite{wang2019physics}, which encompasses an autoencoder based on LSTM and uses $\mathbf{z}_t$ as the input and $L_{static}$ as the loss function. We refer to this model as LSTM$_{ref}$. %The network structure of LSTM$_{ref}$ is the same as CHIMERA except for the inputs and the loss functions. The authors demonstrated the effectiveness of $L_{static}$ towards defending FDIAs, but they did not investigate whether the estimated states would affect the results of contingency analysis in a further step.

The MLP and LSTM$_{ref}$ are trained by following the same procedure as CHIMERA, i.e., 70\% of the data is used for training, 15\% for validation, and 15\% for testing. The training times of the three models, deployed on a computing platform with an  NVIDIA GTX 2048 and an eight-core Intel(R) Xeon(R) CPU of 2.60 GHz, are summarized in Table~\ref{tab:train_time}. %\yaodan{We deploy the models on a computing platform with a NVIDIA GTX 2048 and an eight-core Intel(R) Xeon(R) CPU of 2.60 GHz.} 
%and examine its performance on the IEEE 14 bus system with New York independent system operator (NYISO) load data (9030 epochs in total)
Because of the simple network architecture and loss function, the baseline MLP is trained faster than the other two models with a total time consumed for training to be 358.75s. CHIMERA makes a trade-off between training speed and security guarantees. It is trained slower than the other models, i.e., 1191.96s,  because additional computations are conducted in the calculation of loss functions.

\begin{table}[t]
    \centering
    \caption{Training time of the baseline MLP, LSTM$_{ref}$, and CHIMERA.}
    \vspace{-1mm}
    \begin{tabular}{||c|c|c|c||}
        \hline  \hline 
        Model & Coarse train (s) & Fine tune (s) & Total time (s) \\
        \hline
        MLP & 101.56 & 257.19 & 358.75 \\
        \hline 
        LSTM$_{ref}$ & 218.48 & 889.9 & 1108.38 \\
        \hline
        CHIMERA & 233.09 & 958.87 & 1191.96 \\
        \hline \hline 
    \end{tabular}
    \vspace{-2mm}
    \label{tab:train_time}
\end{table}

\subsubsection{Attack Setup}
We select the measurements to attack based on the criticality of buses calculated according to \cite{liu2017recognition}. The buses 1, 2, 3, 4, 5 have the highest criticality. Thus, the meters on those buses are selected in order for the active power measurements to be injected with errors. The optimal attack vector of Eq.~\eqref{eq:eq1} - \eqref{eq:con6} is solved by the Adam algorithm with a learning rate of \num{1e-2}. The attack vector is generated for the measurement vector at each epoch. We observe that more than 99\% of the estimation result residuals from the three models are smaller than $0.5$. Thus, the threshold of $J(\mathbf{\hat{x}})$ is set as $\tau = 0.5$ in the attack model. We run the attacks for different values of $f_{m}$ and select $f_{m} = 3$ based on the magnitudes of the injected errors. The injected errors have similar magnitudes for all three models. For each targeted measurement, the injected errors result in a Mean Absolute Error (MAE) of $0.55$ for the baseline MLP, $0.54$ for LSTM$_{ref}$, and $0.54$ for CHIMERA.

\iffalse
\begin{table}[]
    \centering
     \caption{The power lines of the IEEE 14 bus system and their corresponding flow limits.}
      \resizebox{0.5\textwidth}{15mm}{
    \begin{tabular}{||c|c|c|c|c|c|c|c|c|c|c||}
        \hline \hline
        Line $k$ & 1 & 2 & 3 & 4 & 5 & 6 & 7 & 8 & 9 & 10 \\
        \hline
        From bus & 1 & 1 & 2 & 2 & 2 & 3 & 4 & 4 & 4 & 5\\
        \hline
        To bus & 2 & 5 & 3 & 4 & 5 & 4 & 5 & 7 & 9 & 6\\
        \hline
        $f^{limit}_k$ (MW) & 120 & 65 & 36 & 65 & 50 & 63 & 45 & 55 & 32 & 45\\
        \hline
        \hline
        Line $k$ & 11 & 12 & 13 & 14 & 15 & 16 & 17 & 18 & 19 & 20 \\
        \hline
        From bus & 6 & 6 & 6 & 7 & 7 & 9 & 9 & 10 & 12 & 13\\
        \hline
        To bus & 11 & 12 & 13 & 8 & 9 & 10 & 14 & 11 & 13 & 14\\
        \hline
        $f^{limit}_k$ (MW) & 18 & 32 & 32 & 32 & 32 & 32 & 18 & 12 & 12 & 12\\
        \hline \hline
    \end{tabular}
   }
    \label{tab:flimit}
\end{table}
\fi

\vspace{-1mm}
\subsection{Evaluation of the Estimation Results without Attacks}
\subsubsection{Estimation Accuracy}
\begin{figure}[t]
    \centering
    \includegraphics[width=0.4\textwidth]{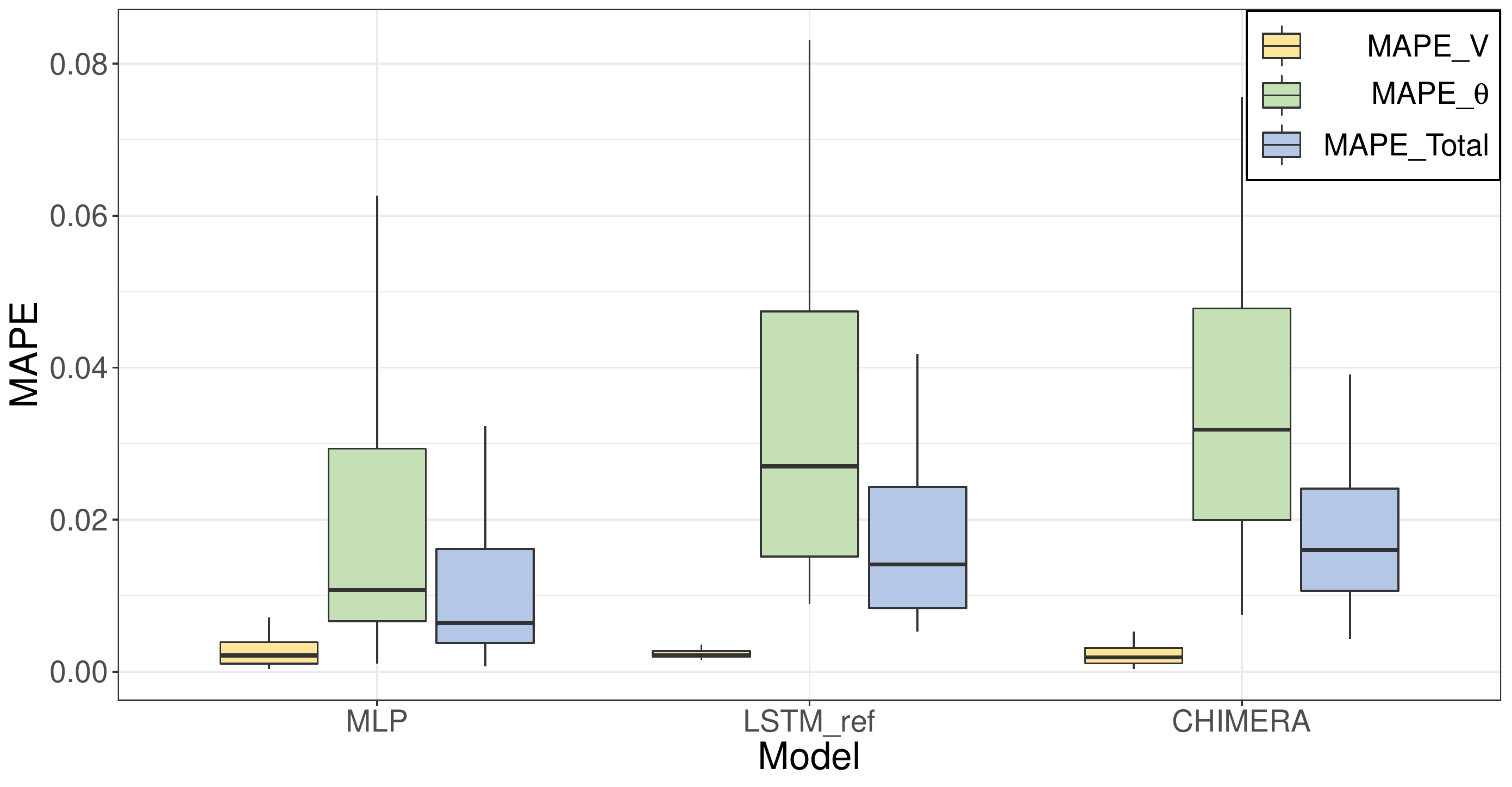}
    \vspace{-1mm}
    \caption{MAPE of the estimated states from the baseline MLP, LSTM$_{ref}$, and CHIMERA in the attack-free case.} %{In the attack-free scenario, MLP achieves the best accuracy in the voltage angle estimations and LSTM$_{ref}$ achieves the best accuracy in the voltage amplitude estimations.}}
    \label{fig:mse_normal}
\end{figure}

Denote the vectors of true states and  estimated states at epoch $t$ as $\mathbf{x}_{t} = [\bm{\theta}_{t}; \mathbf{V}_{t}]$, and $\hat{\mathbf{x}}_t = [\hat{\bm{\theta}}_t;\hat{\mathbf{V}}_t]$, respectively. Here $\bm{\theta}_{t}$, $\hat{\bm{\theta}}_t$, $\mathbf{V}_{t}$ and $\hat{\mathbf{V}}_t$ are the vectors of the true/estimated angles and magnitudes, respectively.
We use the Mean Absolute Percentage Error (MAPE): 
\vspace{-1mm}
\begin{equation}
    MAPE(\mathbf{x}, \mathbf{y}) = \frac{1}{n} \sum_{i=1}^{n} \left|\frac{y_{i}-x_{i}}{ x_{i}} \right|,
\end{equation}
as the accuracy evaluation metric of the estimated states, and define $MAPE\_\theta = MAPE(\bm{\theta}_{t}, \hat{\bm{\theta}}_t)$, $MAPE\_V = MAPE(\mathbf{V}_{t}, \hat{\mathbf{V}}_t)$ and  $MAPE\_Total = MAPE(\mathbf{x}_{t}, \hat{\mathbf{x}}_t)$. The results are summarized in Fig.~\ref{fig:mse_normal}. Since the voltage angles fluctuate greater than the voltage magnitudes, the MSE of the angle estimations are larger than the MSE of the magnitude estimations. All three models achieve satisfiable accuracy with the average MAPEs of the states to be 1.02\% for the baseline MLP, 1.70\% for LSTM$_{ref}$, and 1.76\% for CHIMERA.

\subsubsection{Contingency Analysis Results}

Given the estimated states from the three  models, we perform contingency analysis to reveal the variance of the numbers of $N-1$ and $N-2$ contingencies in the system. By plugging the estimated states $\hat{\mathbf{x}}_t$ into Eq.~(\ref{Eq:measurement}), we obtain the power flows $\hat{\mathbf{f}}_t$. The number of the $N-1$ and the $N-2$ contingencies at epoch $t$ given the estimated power flows $\hat{\mathbf{f}}_t$ are denoted as $\hat{N}_{1, t}$ and $\hat{N}_{2, t}$, respectively. Moreover, the contingency analysis based on the system measurements $\mathbf{z}_t$ at each epoch $t$ is executed to obtain the exact numbers of $N-1$ and $N-2$ contingencies in the system, which are denoted as ${N_{1,t}}$ and ${N_{2,t}}$, respectively. ${N_{1,t}}$ and ${N_{2,t}}$ are referred to as the ground truth. 

\begin{figure} \label{fig:mae_normal}\centering    
\subfigure[] { \label{fig:mae_normalN1}     
\includegraphics[width=0.23\textwidth]{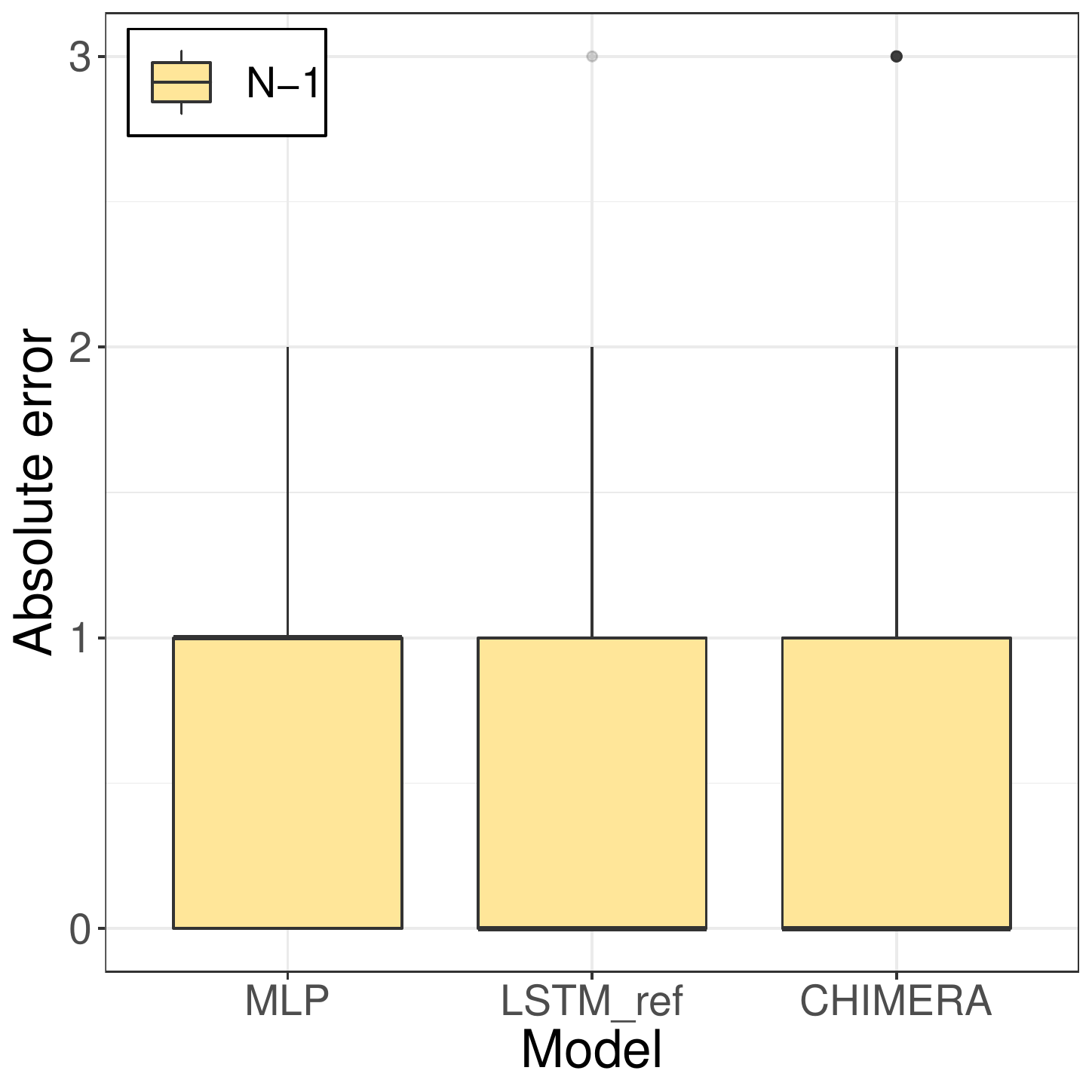}  
 }
% \hfill
\subfigure[] { \label{fig:mae_normalN2}     
\includegraphics[width=0.23\textwidth]{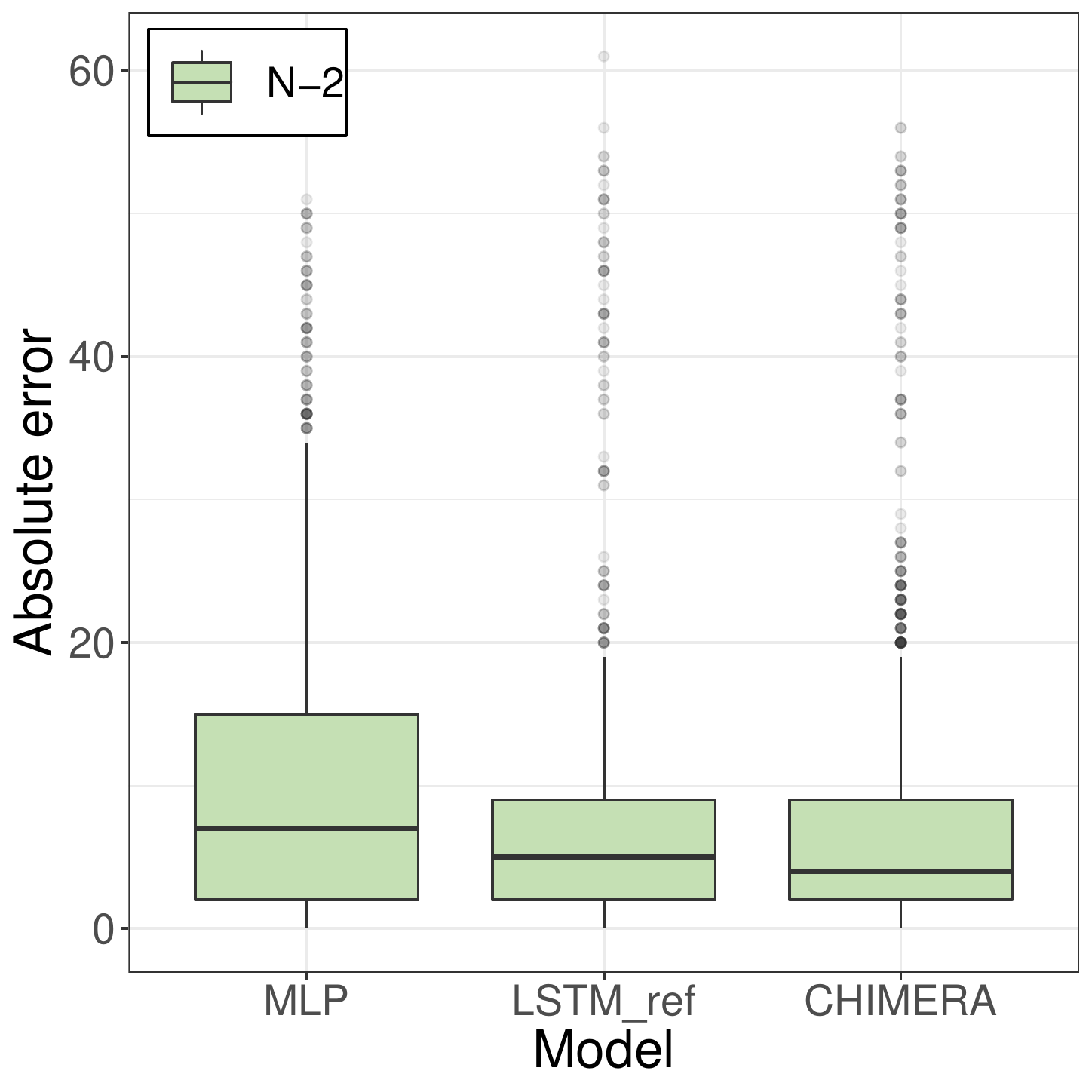}}
%\vspace{-3mm}
\caption{The absolute errors of the numbers of (a) $N-1$ and (b) $N-2$ contingencies given the estimated power flows in the attack-free case.} %{In the attack-free case, LSTM$_{ref}$ and CHIMERA outperform MLP and achieve comparable performance in estimating the number of contingencies given the state estimates from the models.}}
\vspace{-3mm}
\label{mae_normal}
\end{figure}

We use the absolute errors between the aforementioned methods of acquiring the contingency data, indicated with $\epsilon_{1} = |\hat{N}_{1,t} - {N_{1,t}}|$ and $\epsilon_2 = |\hat{N}_{2,t} - {N_{2,t}}|$, as the metric to evaluate the performance of the three models in the attack-free case. The results are shown in Fig.~\ref{mae_normal}. Because of the estimation errors, errors are introduced into the contingency analysis results inevitably. The results demonstrate the benefit of $L_{static}$ over $L_0$. Although the baseline MLP has the smallest MSE of state estimations, LSTM$_{ref}$ and CHIMERA achieve better performance because $L_{static}$ can enforce the consistency between the estimated power flows and the system measurements. For $N-1$ analysis, 68.60\% and 69.12\% of $\hat{N}_{1,t}$ are accurately calculated ($\epsilon_1 = 0$) for LSTM$_{ref}$ and CHIMERA, respectively, while for the baseline MLP, only 46.50\% of $\hat{N}_{1,t}$ are accurately calculated. Besides, for $N-2$ analysis  $\hat{N}_{2,t}$,  the average $\epsilon_{2}$ equals to 7.14 and 7.80 from LSTM$_{ref}$ and CHIMERA, respectively,  while the average $\epsilon_{2}$ for $\hat{N}_{2,t}$ from the baseline MLP is 10.30.

% As shown in Fig. \ref{fig:mae_normal}, the MAE variations of the numbers of $N-1$ and $N-2$ without attacks are compared with the baseline and the LSTM$_{ref}$. More specifically, we could observe that our proposed model has a lower MAE in the $N-1$ scenario comparing with other two models. In the $N-2$ case, it could obtain the number of contingency within a average of $8...$ MAE comparing with the ground truth. 

\vspace{-2mm}
\subsection{Impact of False Data Injection Attacks on Contingencies}
\vspace{-1mm}
The performance of the three models against FDIAs is summarized in Table~\ref{tab:res_sum}. Overall, LSTM${_{ref}}$ and CHIMERA achieve better performance compared with the baseline MLP. Regarding the impacts of FDIAs on $N-2$ contingencies, CHIMERA shows higher resilience  compared to  LSTM$_{ref}$. %Detailed evaluations and analyses are presented in the following.

\begin{table}[t]
\vspace{-1mm}
    \centering
    \caption{Average performance of the three models against FDIAs.}
    \label{tab:res_sum}
    \begin{tabular}{||c|c|c|c|c|c||}
    \hline
    \hline
        Model & MAPE\_V & MAPE\_$\theta$ & MAPE\_Total & $\epsilon_1^\mathbf{a}$ & $\epsilon_2^\mathbf{a}$\\
        \hline
        MLP &  0.27\% & 0.84\% & 0.54\% & 0.35 & 9.16 \\
        \hline
        LSTM${_{ref}}$ & 0.007\% & 0.12\% & 0.06\% & 0.03 & 5.75 \\
        \hline
        CHIMERA & 0.008\% & 0.14\% & 0.07\% & 0.06 & 1.70 \\
        \hline
        \hline
    \end{tabular}
    \vspace{-4mm}
\end{table}

\subsubsection{Estimation Accuracy}
\begin{figure}[t]
    \centering
    \includegraphics[width=0.4\textwidth]{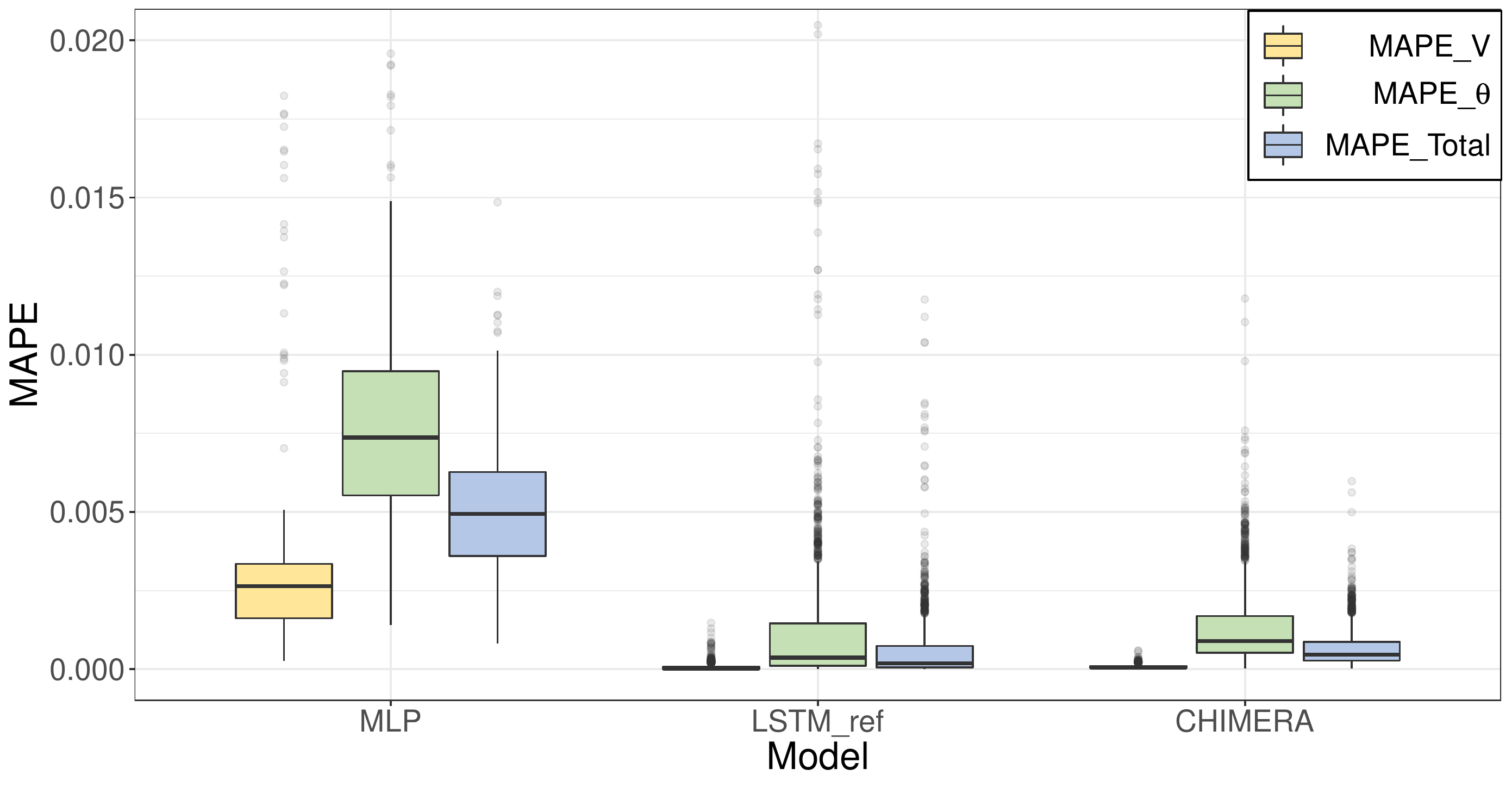}
    \vspace{-1mm}
    \caption{The impact of the attacks on the estimation accuracy.} %{In the occurrence of FDIAs, LSTM$_{ref}$ and CHIMERA outperform MLP and achieve comparable performance regarding the state estimation accuracy.}}
    \vspace{-4mm}
    \label{fig:mse_attack}
\end{figure}

Denote the estimated states from attacked measurements as $\hat{\mathbf{x}}^\mathbf{a}_t$. The impact of the attacks on the estimated states is assessed based on the $MAPE\_\theta^\mathbf{a} = MAPE(\hat{\bm{\theta}}_t, \hat{\bm{\theta}}^\mathbf{a}_t)$, $MAPE\_V^\mathbf{a} = MAPE(\hat{\mathbf{V}}_t, \hat{\mathbf{V}}^\mathbf{a}_t)$ and  $MAPE\_Total^\mathbf{a} = MAPE(\hat{\mathbf{x}}_t, \hat{\mathbf{x}}^\mathbf{a}_t)$. The results are summarized in Fig.~\ref{fig:mse_attack}. 
Note that the attacker intends to affect the contingencies while remaining undetected from the state estimation. 
The results verify the stealthiness of our attack model. We observe that the attacks do not induce large errors to the estimated states: the changes in the estimated states are only 0.54\%, 0.06\% and 0.07\% for the baseline MLP, LSTM$_{ref}$, and CHIMERA, respectively. Despite the slight distinctions, we can still conclude that LSTM$_{ref}$ and CHIMERA are more resilient to FDIAs due to their network architecture and the usage of $L_{static}$.

\subsubsection{Contingency Analysis Results}
\begin{figure} \label{fig:mae_attack}\centering    
\subfigure[] { \label{fig:mae_attackN1}     
\includegraphics[width=0.23\textwidth]{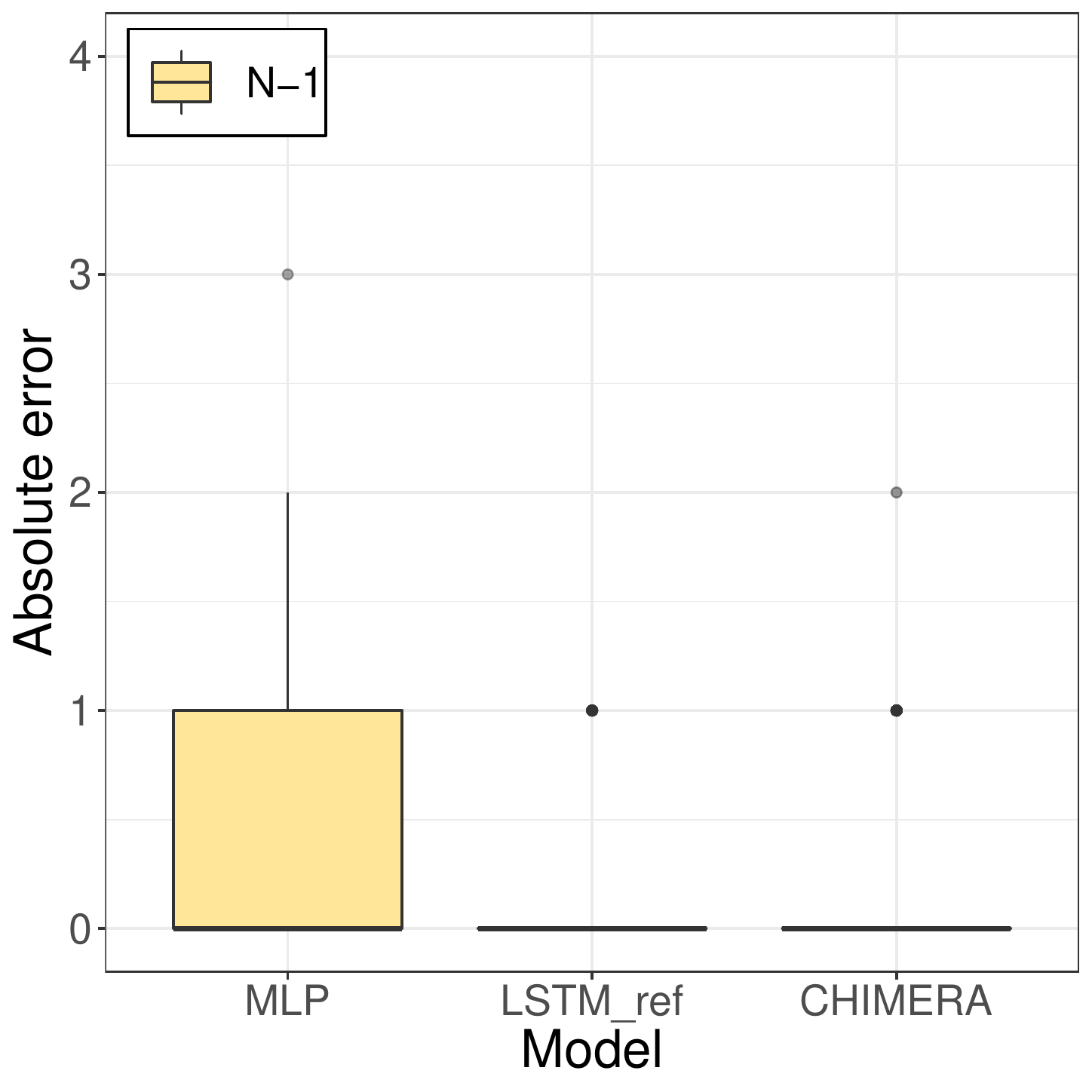}  
 }
% \hfill
\subfigure[] { \label{fig:mae_attackN2}     
\includegraphics[width=0.23\textwidth]{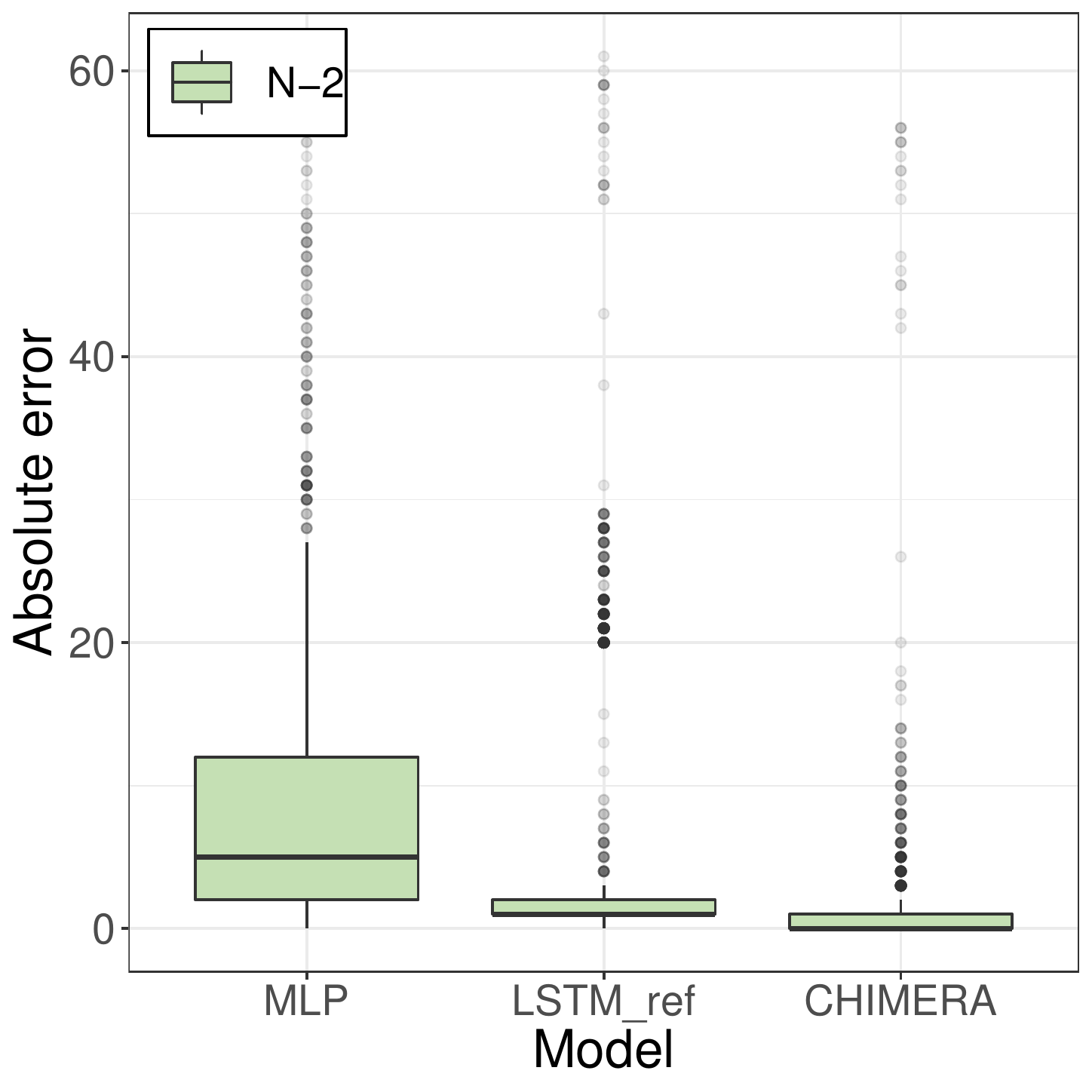}}
\vspace{-4mm}
\caption{The attack impact on (a) $N-1$ and (b) $N-2$ contingency analysis.} % {During occurrence of FDIAs, LSTM$_{ref}$ and CHIMERA outperform MLP and achieve comparable performance regarding $N-1$ contingencies. CHIMERA outperforms LSTM$_{ref}$ regarding $N-2$ contingencies.}}
\label{mae_attack}
\vspace{-5mm}
\end{figure}

Denote the number of the $N-1$ and the $N-2$ contingencies at epoch $t$ given the power flows estimated from the attacked measurements as $\hat{N}^\mathbf{a}_{1,t}$ and $\hat{N}^\mathbf{a}_{2,t}$.
To assess the impacts of the attacks on the contingency analysis, we use the absolute errors between the number of contingencies from estimated power flows before and after attacks, i.e., $\epsilon^\mathbf{a}_{1} = |\hat{N}^\mathbf{a}_{1,t} - \hat{N}_{1,t}|$ and $\epsilon^\mathbf{a}_2 = |\hat{N}^\mathbf{a}_{2,t} - \hat{N}_{2,t}|$, as the performance metrics. The results are presented in Fig.~\ref{mae_attack}. If $\epsilon^\mathbf{a}_1$ or $\epsilon^\mathbf{a}_2$ are not equal to 0, an attack is considered successful. Besides, the larger $\epsilon^\mathbf{a}_1$ or $\epsilon^\mathbf{a}_2$ are, the larger the impact of the attack is. We observe that the contingency analysis results are  sensitive to the accuracy of the estimated states. Although the injected attack vectors have similar magnitudes and only slightly affect the accuracy of the estimated states, the impacts of the attacks on the contingency analysis results from the three models differ a lot. Since no defense is embedded in the baseline MLP, the performance of the baseline MLP is heavily degraded. In the $N-1$ case, 53.50\% of $\hat{N}^\mathbf{a}_1$ are changed ($\epsilon^\mathbf{a}_1 \not = 0$) for the baseline MLP, while the percentage of $\hat{N}^\mathbf{a}_{1}$ changed for LSTM$_{ref}$ and CHIMERA are only 31.4\% and 22.69\%, respectively. 
The maximum $\epsilon^\mathbf{a}_{1}$ is 4 for the baseline MLP, while it is 1 and 2 for LSTM$_{ref}$ and CHIMERA, respectively. In the $N-2$ case, the average $\epsilon^\mathbf{a}_{2,t}$ is 9.16 for the baseline MLP, while the average $\epsilon^\mathbf{a}_{2,t}$ is 5.75 for LSTM$_{ref}$ and 1.70 for CHIMERA. Moreover, the results from LSTM$_{ref}$ show that using only  $L_{static}$ cannot totally defend against FDIAs. On the other hand, because of the usage of the $L_{dynamic}$, the impact of FDIAs on CHIMERA is significantly limited. Specifically, 64.81\% of attacks fail to take effect on CHIMERA, i.e., $\epsilon^\mathbf{a}_2 = 0$, while the percentages for the baseline MLP and LSTM$_{ref}$ are only 7.14\% and 22.32\%, respectively. Moreover, 91.74\% of attacks have limited impacts on CHIMERA, i.e., $\epsilon^\mathbf{a}_2 < 5$, while for the  baseline MLP and LSTM$_{ref}$ these values are 48.36\% and 79.32\%, respectively. 

{\vspace{-1mm}
\subsection{Practical Implications and Applications}
\vspace{-1mm}
In terms of real-world applications, CHIMERA can be implemented at the computing stations of power grid operators and be part of the EMS. For example, it can be deployed as an additional application in the EMS by updating the existing state estimation routines. Thus, CHIMERA does not require or induce any hardware modifications or overhead. The major computation cost of CHIMERA is on the training process. Despite that CHIMERA requires longer training time, it can be trained offline abd it  does not induce additional computational overhead during runtime. In fact, the times for CHIMERA and MLP/LSTM$_{ref}$ to estimate states are of the same order and approximately 0.05ms, which are neglectable and do not violate any real-time requirements \cite{8624411}. Furthermore, during attacks, significant enhancement has been achieved by CHIMERA in estimating the number of $N-2$ contingencies. {For the IEEE 14-bus system, there can be 190 $N-2$ contingencies in total. Through our experiments, we show that in 91.74\% attacks, CHIMERA can achieve an estimation accuracy more than 97.4\% (i.e., $\epsilon^\mathbf{a}_2 < 5$) for $N-2$ contingencies.} With such high accuracy, CHIMERA guarantees the normal operation of the power grid during the occurrence of FDIAs.

}

% Take the consideration of the worse case scenario, there can be 190 $N-2$ contingencies in the IEEE 14-bus system, the estimation accuracy of CHIMERA increase to 97.4\%. 

% For the IEEE 14-bus system, there can be 190 $N-2$ contingencies in total. Through our experiments, we show that in 91.74\% attacks, CHIMERA can achieve an estimation accuracy more than 97.4\% (i.e., $\epsilon^\mathbf{a}_2 < 5$) for $N-2$ contingencies.

% \begin{figure}[!htbp]
%     \centering
%     \includegraphics[width = 0.42\textwidth]{Fig/MAEAttackS.pdf}
%     \caption{The impacts of the attacks on the contingency analysis.}
%     \label{fig:mae_attack}
% \end{figure}

% \begin{figure}[!htbp]
%     \centering
%     \includegraphics[width = 0.47\textwidth]{Fig/com/proposed1.pdf}
%     \caption{.}
%     \label{fig:proposed}
% \end{figure}

% \begin{figure*} \centering    
% \subfigure[] { \label{fig:1}     
% \includegraphics[width=5.7cm]{Fig/com//MLP2.pdf}  
%  }
% % \hfill
% \subfigure[] { \label{fig:2}     
% \includegraphics[width=5.7cm]{Fig/com/lstm2.pdf}}
% % \hfill
% \subfigure[] { \label{fig:3}     
% \includegraphics[width=5.7cm]{Fig/com/proposed2.pdf}}
% % \hfill }

% \caption{ \subref{fig:1}, \subref{fig:2}, \subref{fig:3}} 
% \label{fig:resultsattacks}  
% \end{figure*}
\vspace{-1mm}
\section{Conclusions}\label{s:conclusion}\label{s:6}
In this paper, we %investigate the impacts of FDIAs on contingency analysis. %We observe that the contingency analysis results is pretty sensitive to the state estimation results, and it is insufficient to evaluate the performance of estimators with only state estimation accuracy. 
investigate an attack model intending to disturb power systems contingencies through FDIAs. We show that the attack can manipulate contingency analysis accuracy by slightly increasing the state estimation errors. To mitigate the effects, we propose CHIMERA, a hybrid attack-resilient estimator which ensures the accuracy of state estimation and the resulting contingency analysis. CHIMERA leverages the dynamic and static features of the power grid observation model and embeds them into a deep learning model. %Simulations are conducted on a power grid benchmark case with synthetic data generated from NYISO load data. Our results justify the resilience of CHIMERA to FDIAs and the capability of maintaining system operations in the EMS functions of state estimation and contingency analysis. 
\vspace{-2mm}
\bibliographystyle{IEEEtran}
\bibliography{biblio}

\end{document}